# Quand le jeu vidéo est le catalyseur d'expérimentations théâtrales (2014-19)
Georges Gagneré, metteur en scène & Cédric Plessiet, artiste numérique

## 1. Découverte de la marionnette virtuelle (2014-2015)
### 1.1. Une rencontre décisive

Metteur en scène professionnel et familier du travail avec les technologies numériques audiovisuelles[1], Georges Gagneré a été choisi en septembre 2013 par le département théâtre de l'UFR Art de l'Université Paris 8 pour enseigner comment intégrer les médias audiovisuels dans la dramaturgie, les techniques de jeu d'acteur et plus largement la pratique scénique. Début septembre, le metteur en scène prend aussitôt rendez-vous avec la directrice de l'équipe de recherche Image Numérique et Réalité Virtuelle (INREV), dont les salles de travail sont situées à dix mètres du Studio-Théâtre dans lequel il s'apprête à commencer ses cours à la fin du mois. Souhaitant connaître les travaux d'un enseignant-chercheur sur le cinéma d'animation, Marie-Hélène Tramus lui présente cependant un autre enseignant chercheur de son équipe, Cédric Plessiet, dont le domaine de recherche principal est l'acteur virtuel. Lors de cette première rencontre, ce dernier évoque une de ses dernières recherche-créations, *Lucky*, réalisée l'année précédente en 2012 : une installation interactive d'art numérique inspirée de *En attendant Godot*, de Samuel Beckett, dans laquelle le spectateur est invité à porter un casque de réalité virtuelle et à s'immerger dans le corps du personnage Pozzo afin de jouer le rôle du maître capricieux qui tient en laisse et martyrise son serviteur Lucky, joué par un acteur virtuel réalisé d'après le scan 3D d'un danseur professionnel, et doté d'un comportement partiellement autonome. Le spectateur se voit tenir une laisse virtuelle, qu'on lui a glissé parallèlement dans la main, et avec laquelle il tient Lucky par le cou. Lorsqu'il tire sur la corde, Lucky résiste, mais il peut aussi tomber à terre. L'interaction est réalisée avec un bras à retour d'effort. Lucky obéit aussi aux ordres simples qu'on lui crie : avance, arrête, etc.

L'artiste numérique indique par ailleurs que l'installation résulte d'un ambitieux projet de recherche en cours de finalisation : OUTILNUM[2], plate-forme de développement dédiée à la prévisualisation temps réel en situation pour le cinéma et le jeu vidéo (cf. fig. 3 gauche). Le metteur en scène, fortement impressionné par ces résultats et par leur relation au théâtre et à l'acteur, indique qu'il souhaite sensibiliser ses futurs étudiants de théâtre à l'art numérique interactif, et en particulier aux acteurs virtuels dont l'expressivité scénique dans certains jeux vidéo est stupéfiante. Marie-Hélène Tramus rappelle alors que les départements Arts et Technologies de l'Image numérique (ATI) et Théâtre ont déjà eu l'occasion de collaborer, et que dans l'éco système de recherche et d'enseignement propre à Paris 8, fortement transdisciplinaire et favorisant les pratiques de recherche-création, il est tout à fait souhaitable que cela continue. Cette rencontre inattendue entre deux artistes et chercheurs travaillant chacun étroitement avec des acteurs, soit numériques soit physiques, marque le démarrage d'un dialogue fructueux entre théâtre et jeu vidéo. A l'issue du rendez-vous, l'artiste numérique invite le metteur en scène à participer en fin de semestre à un atelier IDEFI

---

[1] Georges Gagneré, « Émergence et fragilité d'une recherche-création (2000-2007) », *Ligeia, dossier sur l'art*, « Théâtres Laboratoires. Recherche-création et technologies dans le théâtre d'aujourd'hui », janvier-juin 2015, Paris, p. 148-158.

[2] Cédric Plessiet et al., « Autonomous and interactive virtual actor, cooperative virtual environment for immersive Previsualisation tool oriented to movies », *in Proceedings of the 2015 Virtual Reality International Conference (VRIC '15)*, ACM, 2015, New York, USA.

CréaTIC[3] qu'il organise avec des étudiants en art, intitulé « Du geste capté au geste d'interactivité numérique ».

**1.2. Découverte d'entités virtuelles en action**

En janvier 2014, Georges Gagneré assiste alors au déroulement de cet atelier intensif, qui décline la plate-forme OUTILNUM et sa librairie de programmation AKeNe au service d'expérimentations artistiques conduites par des artistes en danse, art numérique et réalisation sonore. Une danseuse, équipée d'une combinaison de capture de mouvement Optitrack, improvise en immersion dans un environnement virtuel en générant l'environnement sonore et graphique par la danse d'une marionnette virtuelle qu'elle contrôle par son propre corps. Le metteur en scène expérimente aussi l'installation *Lucky* qui a été remise en place à l'occasion du stage et rencontre son premier golem virtuel (cf. 4.1.). OUTILNUM est donc une plate-forme d'expérimentation bien concrète et accessible aux étudiants. Elle fonctionne avec Optitrack, mais aussi avec un périphérique beaucoup plus commode d'utilisation : la caméra Kinect de Microsoft. Il demande alors s'il serait envisageable de faire une démonstration du dispositif dans le cadre d'un prochain cours de la licence théâtre. C'est ainsi que les étudiants de théâtre sont invités en mai 2014 au studio-théâtre à découvrir *Lucky* et à rencontrer en immersion une marionnette virtuelle qu'ils animent eux-mêmes en improvisant devant une caméra Kinect. La dimension ludique du dispositif facilite la sensibilisation à l'utilisation des technologies numériques sur la scène, d'habitude rétive à toute intrusion informatique.

Le metteur en scène obtient alors un financement de l'université pour acquérir un ordinateur portable suffisamment puissant pour supporter un casque de réalité virtuelle et construire au sein du département théâtre une plate-forme d'expérimentation de la réalité virtuelle et des outils du jeu vidéo dans une perspective théâtrale pour ses étudiants : la Station d'Actions Scéniques en 3D (SAS3D) voit le jour. Elle sera directement utilisée à l'occasion d'un renouvellement de l'atelier CréaTIC en janvier 2015, dans lequel le metteur en scène est invité officiellement à coconstruire les expérimentations avec des étudiants issus d'art numérique, de cinéma et de théâtre. A cette occasion, l'artiste numérique adaptera OUTILNUM pour réaliser un dispositif mettant en relation deux performeurs, chacun équipé d'un casque de réalité virtuelle et contrôlant avec une Kinect un avatar stylisé (de feu ou de glace). Réalité virtuelle et réalité mixte coexistent dans l'improvisation des deux acteurs physiques qui évoluent de part et d'autre d'un écran, sur lequel est projeté pour le public l'espace virtuel commun où ils interagissent avec leurs marionnettes virtuelles (cf. fig. 1 gauche). Le dispositif permettra ensuite de sensibiliser les étudiants de théâtre à l'habitation d'une marionnette virtuelle et aux contraintes de jeu en immersion[4]. C'est un premier prototype dédié à la pratique théâtrale et directement issu de la plate-forme OUTILNUM et de la librairie AKeNe. En parallèle de cette exploration pratique, les laboratoires INREV (EA1410) et Scène du monde, création, savoirs critiques (EA 1573) lancent au 1er semestre 2015 une série de six ateliers-rencontres pour démarrer un état de l'art sur le thème *Corps humain, avatar numérique et arts vivants*[5]. C'est

---

[3] Dans le cadre des Initiatives d'Excellence en Formations Innovantes financées par l'Agence Nationale de la Recherche, le programme IDEFI CréaTIC démarre à Paris 8 en 2013-14.

[4] Georges Gagneré et Cédric Plessiet, « Traversées des frontières », *in Frontières numériques & artéfacts*, dir. Hakim Hachour & al., L'Harmattan, 2015, p. 9-35.

[5] Coorganisés par Véronique Muscianisi, Georges Gagneré et Cédric Plessiet, les ateliers ont permis les rencontres suivantes: « Acteurs virtuels, corporéité et spectacle vivant » (le 18 décembre 2014 avec J.-F. Jégo, M. Bret et M. Passedouet), « Émotions et effets de présence pour le réalisme des avatars » (le 12 février 2015 avec E. Grimaud, G. Khemiri, R. Ronfard), « Augmentation versus disparition de l'acteur à travers le prisme du numérique » (le 5 mars

l'occasion de croiser des points de vue complémentaires sur l'acteur, qui consolident l'intuition que les domaines du théâtre et du jeu vidéo peuvent dialoguer ensemble[6].

**1.3. Un premier déplacement du jeu vidéo vers le théâtre**
L'intensification de ce dialogue entre jeu vidéo et théâtre va induire un changement d'outil. En mars 2015, à la suite d'un changement de modèle économique, la société américaine Epic Game rend son moteur de jeu vidéo Unreal Engine 4 (UE4) utilisable gratuitement pour les activités d'enseignement et de recherche, alors que l'autre moteur très répandu chez les développeurs indépendants, Unity 3D, devient payant. Par ailleurs, grâce à son formalisme Blueprint de programmation graphique en code C++, UE4 est accessible aux artistes sans formation de codeur. Les gens de théâtre pourraient donc se familiariser avec les techniques de développement des jeux vidéo sans avoir besoin de se former à la pratique ardue du codage en C++. Ces deux arguments convainquent Cédric Plessiet de basculer l'implémentation d'OUTILNUM de Unity 3D vers UE4 et d'impliquer directement Georges Gagneré dans la construction des outils. Le metteur en scène souhaite proposer aux étudiants et chercheurs en théâtre un moyen de s'approprier le corps des avatars en approfondissant les contraintes de manipulation liées à la caméra Kinect V1, seule version utilisable dans le système informatique sous Windows 7 de l'université Paris 8. Le serveur AKeNe qui récupère les informations de la Kinect V1 est adapté à la version 4.8 d'UE4 et un avatar « filaire » est construit sous forme de blueprint à partir des articulations repérées par la Kinect.

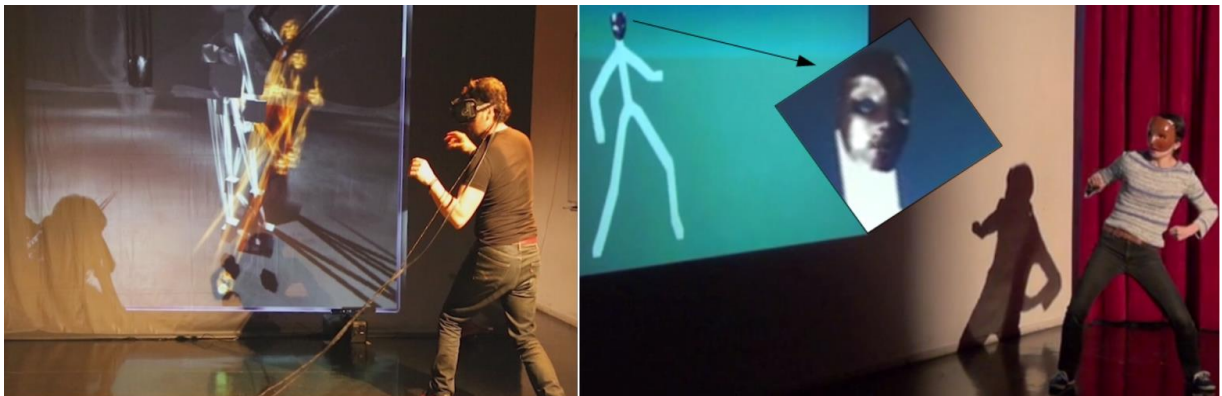

*Figure 1. Gauche : IDEFI CréaTIC de janvier 2015. Droite : CW#1 en décembre 2015.*

Cette transposition d'OUTILNUM sur UE 4.8 permet alors d'envisager deux projets. Le premier constitue le volet « Masque et technologies » du projet de recherche *La Scène Augmentée*, porté par Erica Magris à partir de 2015 au sein du LABEX ARTS H2H[7] de l'université Paris 8. La proposition consiste à mettre en dialogue la technique de jeu avec masque et les nouvelles technologies, et notamment la figure de l'avatar. Il s'agit d' « explorer le potentiel d'expressivité et d'interaction entre spectateurs, acteurs, personnages masqués et

---

avec C. Bardiot et J.-F. Ballay), « L'interaction entre humain et avatar » (le 26 mars avec M. Courgeon, C. Jost, B. Andrieu), « La direction d'acteurs et d'avatars » (le 16 avril avec Y. Tina, G. Gagneré, J. Guez, J.-F. Jégo et C. Plessiet) et « Avatar et mise en scène » (le 28 mai avec J. Huang, A. Dubos, B. Cheval).

[6] Georges Gagneré et Cédric Plessiet, « Échanges entre metteur en scène et artiste numérique à propos de la direction d'« acteur » », *in La direction d'acteurs peut-elle s'apprendre ?*, dir. Jean-François Dusigne, Les Solitaires intempestifs, 2015, p. 360-374.

[7] Le Laboratoire d'excellence des arts et médiations humaines ARTS H2H a été lauréat des Investissements d'avenir en 2011. Il s'est transformé en 2018 en École Universitaire de Recherche EUR ARTEC, qui promeut des dispositifs émergents de formation.

avatars en recourant aux techniques d'improvisation du jeu masqué[8] ». Les 11 et 12 décembre 2015, le premier atelier CW#1 associe fructueusement contrôle d'un avatar avec Kinect et jeu avec masque (cf. fig. 1 droite). Il sera complété par deux autres ateliers sur 2016.

**1.4 Mettre la main à la pâte**
Le second projet sera la réalisation en janvier 2016 d'un troisième et dernier CréaTIC « Du geste capté au geste d'interactivité numérique » mettant en dialogue des étudiants en master des départements Théâtre et ATI. Un enjeu de cet atelier est d'approfondir la qualité des gestes captés avec des comédiens expérimentés en vue de construire des marionnettes virtuelles avec des corps plus élaborés qu'une simple transposition des articulations captées par la Kinect V1. Il s'agit dans un deuxième temps d'utiliser ces actions au sein de deux instruments de contrôle de personnages du domaine de l'intelligence artificielle pour les jeux vidéo : l'automate à états finis et l'arbre à comportement. Le premier instrument enchaîne des actions en fonction de conditions d'exécution, et le second, plus élaboré, permet de définir un comportement en fonction d'une analyse de contexte. Seul l'automate à états finis sera finalement exploité. Il conduit les étudiants à s'approprier une chaîne d'acquisition de gestes en utilisant la Kinect, Motion Builder (Autodesk) et un transfert final vers UE 4.8 en passant par Maya (Autodesk). Les gestes sont ensuite joués à l'aide du clavier d'un ordinateur pour improviser des comportements sur trois marionnettes virtuelles construites par Cédric Plessiet spécifiquement pour l'atelier. La station Optitrack est dédiée à la réalisation d'improvisation entre un comédien manipulant un des personnages, immergé dans une scénographie 3D et en relation avec un des trois autres personnages, contrôlé au clavier par un autre comédien et effectuant les gestes préenregistrés en amont avec la Kinect.

L'utilisation du logiciel Motion Builder conduit à s'approprier le fonctionnement de la marionnette virtuelle. Le dispositif global aborde la complexité à rendre vivante une marionnette animée par des gestes préenregistrés face à une marionnette vivante contrôlée par de la capture de mouvement en temps réel. En retour, la manipulation de l'automate à états finis et les potentialités exposées de l'arbre à comportement permettent d'appréhender l'expressivité d'entités virtuelles développant des comportements autonomes embryonnaires, et intéressants sur le plan de la construction des actions scéniques d'un personnage. La projection de mêmes mouvements sur des avatars d'apparence différente surprend aussi les étudiants de théâtre. Ils retrouvent des questionnements rencontrés sur le précédent CW#1 avec le jeu masqué, notamment l'influence du masque sur la gestualité de l'acteur qui le porte.

**2. Entre immersion et réalité mixte, animation temps réel et préenregistrée (2016)**
**2.1. Première approche de la direction d'avatar**
Dans son propre processus de recherche-création, Georges Gagneré avait commencé à explorer la figure de la marionnette virtuelle au printemps 2015 dans le cadre de deux résidences de création au Laboratoire d'informatique de Bordeaux autour de son projet *ParOral*[9]. Il s'agit de faire interagir des avatars en relation avec la prosodie d'un comédien lisant un conte d'Andersen, *L'Ombre*. Le cœur de la recherche porte sur l'utilisation du séquenceur temporel Iscore[10] pour organiser des événements scéniques qui dépendent de la performance orale du lecteur. Le metteur en scène souhaite manipuler un environnement 3D

---

[8] Cf. Programme du premier atelier « Masque et Technologies : Immersion, expression, interaction – cluster workshop#1 » (CW#1)
[9] http://didascalie.net/rech-paroral (consulté le 22/01/2020)
[10] Iscore est un séquenceur inter-médias libre et open-source pour le codage créatif (https://ossia.io/ consulté le 22/01/2020)

mettant en scène des silhouettes plates à forme humaine (encore nommées ombravatars) et collabore avec un compositeur qui réalise l'environnement sonore à partir d'une transformation de la voix du lecteur. Il sollicite alors Cédric Plessiet pour la construction d'un prototype de manipulation d'ombravatars qui pourraient jouer des actions scéniques préenregistrées. Il s'agit alors d'implémenter sous Unity quatre ombravatars contrôlés chacun par un automate à états finis permettant de combiner une dizaine d'actions préenregistrées, et de relier Unity à Iscore à l'aide du protocole Open Sound Control pour pouvoir en recevoir les instructions. Les actions, qu'il a préenregistrées avec la station Optitrack, permettent au metteur en scène d'écrire des interactions scéniques élémentaires entre les ombravatars et le lecteur au fil de la lecture du conte, et de s'exercer pour la première fois à la direction d'avatar.

Le projet *ParOral* donne lieu à une troisième résidence en avril 2016 en vue de finaliser l'expérimentation avec une nouvelle version du logiciel Iscore. A cette occasion, le metteur en scène utilise alors la nouvelle version du prototype d'expérimentation construit avec UE 4.8 (cf. fig. 2 gauche) et approfondit les techniques acquises à l'occasion de l'atelier CréaTIC de janvier 2016 pour développer un automate à états finis plus élaboré explorant des enchaînements scéniques plus expressifs et plus variés que le précédent dispositif sous Unity, qu'il ne pouvait pas lui-même faire évoluer. Cela lui permet de comprendre concrètement l'importance fondamentale de pouvoir transformer directement les outils numériques pour s'approcher de résultats satisfaisants sur le plan de l'expression scénique, de la même manière qu'un metteur en scène doit connaître le fonctionnement des diverses pratiques théâtrales qui permettent de fabriquer un spectacle pour pouvoir le faire évoluer dans la direction souhaitée.

**2.2. Co-construction du dispositif numérique expérimental**
L'exploration poussée de la gestualité des marionnettes virtuelles incite Cédric Plessiet à porter sur Windows 10 le prototype expérimental développé avec UE 4.8 afin de pouvoir utiliser la Kinect V2 qui a la particularité d'offrir un joint supplémentaire de contrôle au niveau de la tête et de pouvoir directement être utilisé dans UE4 grâce au plugin Kinect4Unreal[11]. Le prototype d'expérimentation devient donc plus précis et moins compliqué à utiliser puisque le logiciel tiers Motion Builder n'est plus nécessaire pour transmettre les mouvements. Cette évolution se concrétise à l'occasion de la conférence-performance Cou2garnak qu'il donne dans le cadre du second atelier CW#2 du projet « Masque et Nouvelles technologies » qui fait se croiser constructeurs de masque et animateurs d'acteurs virtuels le 4 mars 2016. La communauté théâtrale de Paris 8 découvre ainsi pour la première fois une marionnette virtuelle improvisant avec son maître, et qui préfigure l'adaptation de golems virtuels à une utilisation dans un cadre théâtral. Avec OUTILNUM sur UE4 et la Kinect V2, l'artiste numérique met en scène la rébellion de son double en désaccord avec les propos tenus dans la conférence. Il franchit alors un pas supplémentaire vers le théâtre et se fait acteur pour l'occasion. Cette mobilité des points de vue entre théâtre et jeu vidéo est fondamental dans le dialogue entre l'artiste numérique et le metteur en scène. La progression se fait pas à pas en s'appropriant de nouveaux outils et de nouvelles approches expérimentales.

---

[11] Kinect4Unreal a été développé pour les versions UE 4.7 à 4.20 par le groupe Opaque. Le plugin permettait d'utiliser la version 2 de la caméra Kinect sans passer par Motion Builder.

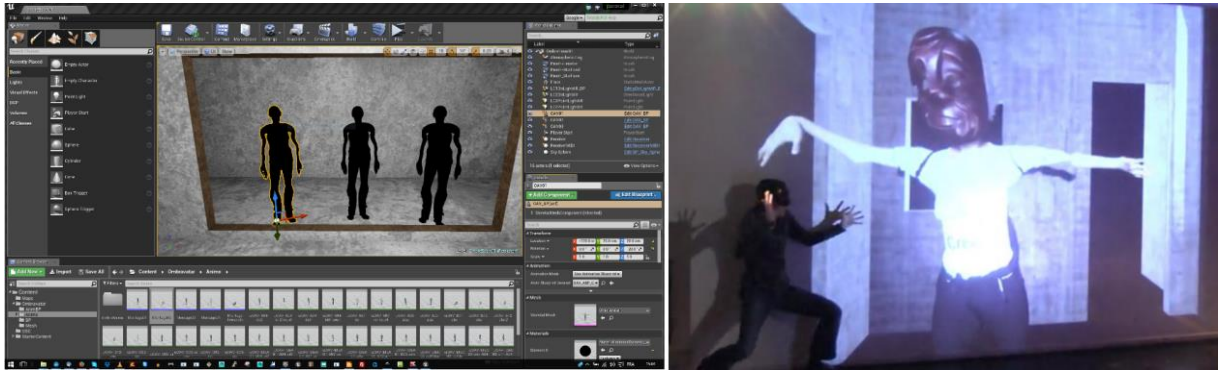

*Figure 2. Gauche :* ParOral *en avril 2016. Droite :* CW#3 *en mai 2016.*

L'évolution d'OUTILNUM vers UE4 et la Kinect V2 offre une première occasion au metteur en scène de participer au développement informatique des outils. Principal opérateur numérique pendant les expérimentations de CW#1, il se charge d'adapter les avatars déjà créés pour les utiliser à l'atelier suivant CW#3 et modifie les blueprints pour intégrer le nouveau comportement de la tête, particulièrement important lorsque l'on expérimente le port du masque sur les nouveaux avatars créés pour l'atelier CréaTIC[12]. En mai 2016, CW#3 permet d'approfondir le dialogue entre des acteurs masqués et des acteurs contrôlant des avatars humanoïdes masqués, dans un dispositif expérimental qui redistribue les rôles entre observation, jeu et construction technologique (cf. fig. 2 droite). En effet, chaque participant est invité à expérimenter la position du spectateur, de l'acteur contrôlant l'avatar et de l'opérateur contrôlant la capture du mouvement et son transfert sur les avatars dans une scénographie 3D. L'exploration théâtrale des outils du jeu vidéo commence par l'assemblage de briques élémentaires dont chacun peut se saisir et qui mettent en relief une nouvelle place possible que les artistes numériques pourraient occuper comme co-auteur d'un dispositif scénique. En effet, le metteur en scène ne peut pas se contenter d'exiger des effets spéciaux visuels. De même que l'artiste numérique s'était fait acteur sur CW#2, le metteur en scène doit comprendre le fonctionnement des nouvelles entités virtuelles, ce qui dessine de nouvelles compétences à acquérir pour les artistes de théâtre[13]. La réussite de CW#3 auprès des chercheurs, des étudiants et des professionnels valide la pertinence d'une exploration théâtrale des possibilités expressives offertes par les arts vidéoludiques.

**2.3. De Kinect à Perception Neuron : accroître l'expressivité des marionnettes virtuelles**
Dans la foulée de CW#2, Cédric Plessiet poursuit le portage intégral sur UE 4.11 de la plate-forme OUTILNUM et de la librairie AKeNe associée. Cela sera finalisé en juin 2016 à l'occasion d'une présentation à des enseignants-chercheurs de l'Ecole Louis Lumière, intéressés par les dispositifs de prévisualisation en situation pour le cinéma. Le contrôle vocal des avatars est implémenté ainsi que la possibilité pour le réalisateur et l'acteur de s'immerger dans le point de vue de chaque personnage avec un casque de réalité virtuelle. Grâce à l'intégration de la Kinect V2, le réalisateur peut aussi voir en immersion la représentation 3D

---

[12] Georges Gagneré et Cédric Plessiet, « Perceptions (théâtrales) de l'augmentation numérique », *in Frontières numériques & artéfacts*, Actes du colloque international Frontières Numériques : Perceptions, Toulon, décembre 2016 ⟨hal-02101604⟩

[13] Georges Gagneré et Cédric Plessiet, « Sur la collaboration d'un metteur en scène et d'un programmeur : des synergies aux hybridations des compétences professionnelles. Entretien par Izabella Pluta », Critiques. Regard sur la technologie dans le spectacle vivant. Carnet en ligne de Theatre in Progress, in Web : http://theatreinprogress.ch/?p=455 , mis en ligne le 20 septembre 2018, Izabella Pluta ©

d'un acteur physique capté par la caméra. Le programme combine les informations de profondeur et de couleur pour reconstituer une vue 3D de l'acteur physique, en temps réel, dénommé « avacteur ». Du point de vue théâtral, cette figure est intéressante car elle permet à un comédien d'être présent dans une scène 3D en tant qu'avatar sans besoin d'être scanné ni de porter une combinaison de capture de mouvement. L'avacteur est cependant contraint par le frustrum de la Kinect qui impose au performeur d'être toujours face à la caméra, ce qui avait limité les actions scéniques au fil des expérimentations CW#1 et CW#3. Les deux collaborateurs décident alors d'explorer une solution économique de capture de mouvement mise sur le marché l'année précédente : la combinaison inertielle Perception Neuron fabriquée par Noitom, utilisable dans UE4 par un plugin associé (similaire au plugin Kinect4Unreal).

En décembre 2016, le metteur en scène acquiert des combinaisons de Perception Neuron par l'intermédiaire de la structure théâtrale professionnelle avec laquelle il collabore, didascalie.net[14], et propose de consolider l'application des résultats de prévisualisation d'OUTILNUM au plateau théâtral, dans le cadre d'un dispositif d'expérimentation en réalité mixte qu'il intitule AvatarStaging, et de l'expérimenter en pratique à l'occasion d'un atelier exploratoire le 10 décembre 2016. Il s'agit de permettre l'improvisation entre, d'un côté, une performeuse équipée de la nouvelle combinaison de capture de mouvement Perception Neuron et d'un casque de réalité virtuelle pour voir l'environnement du point de vue de la marionnette virtuelle qu'elle contrôle, et d'un autre côté, un performeur physique devant un écran représentant la marionnette dans son espace, au pied duquel est installée une Kinect V2. Cette Kinect rend présent à la performeuse en immersion son partenaire de deux manières : sous forme d'une représentation vidéo insérée dans l'espace 3D ou bien sous forme d'avacteur. Le résultat est accessible à des spectateurs qui assistent au dialogue entre le performeur devant la Kinect et son partenaire avatar, en réalité mixte ou en immersion. L'expressivité de la performeuse est démultipliée par la Perception Neuron.

Beaucoup de chemin a été parcouru depuis le premier atelier CréaTIC de janvier 2014. En trois ans, Cédric Plessiet a réussi à construire un dispositif d'expérimentation en temps réel avec des avatars, accessible économiquement et facilement déployable sur une scène de théâtre. Georges Gagneré peut alors intégrer à un cours de licence 3 sur le 1er semestre de l'année universitaire 2016-17 la manipulation de marionnette virtuelles avec la kinect V2, et à un second cours pour des L3 au second semestre, la manipulation avec la Perception Neuron. Parallèlement, les résultats positifs obtenus sur le volet « Masque et Nouvelles Technologies » de *La Scène Augmentée* conduisent la porteuse du projet triennal à consacrer la dernière année à une focalisation sur le dialogue entre acteur masqué et avatar à l'occasion d'une série d'ateliers intitulés « Masques et Avatars », dans la continuité de CW#3, et qui se concluront à l'issue d'un CW#8 sur un colloque international intitulé « Masques Technologiques : altérités hybrides de la scène contemporaine » en décembre 2017.

**3. Consolidation des outils : AvatarStaging et AKN_Regie (2017-2018)**
**3.1. Apprivoiser la liberté de mouvement au sein d'AvatarStaging**
La liberté de mouvement offerte par la combinaison Perception Neuron a un double effet sur le développement d'AvatarStaging. Le premier se manifeste par la focalisation des expériences sur le contrôle des marionnettes virtuelles en temps réel par un performeur, que nous appellerons le mocapteur. Cette nouvelle figure sur un plateau théâtral transfère une qualité de présence à l'avatar suffisamment élevée pour permettre des improvisations avec des acteurs physiques, notamment des acteurs masqués. L'étude des marionnettes contrôlées par automate à états finis avec des animations préenregistrées passe au second plan. Le second effet se traduit par une focalisation sur la dimension réalité mixte des dispositifs au détriment des questions d'immersion. Cela coupe le mocapteur de la possibilité d'établir un contact intime avec l'avatar qu'il contrôle et le contraint à développer des techniques de jeu spécifiques en relation avec des moniteurs de retour permettant de visualiser l'espace virtuel qu'il habite. En revanche, travailler dans la même réalité fluidifie la communication entre toutes les parties prenantes des expérimentations : mocapteur, acteur physique, metteur en scène, artiste numérique. C'est une condition de réussite dans un processus théâtral.

---

[14] http://didascalie.net (consulté le 22/01/2020)

C'est à l'occasion des premiers ateliers de *Masques et Avatars* en mars, mai et juin 2017 qu'apparaissent les verrous techniques à résoudre pour rendre AvatarStaging opérationnel en vue d'une utilisation théâtralement intéressante de la Perception Neuron[15] (cf. fig. 3 droite). Trois chantiers surgissent : il faut pouvoir projeter sur différents personnages virtuels les informations de la Perception Neuron transmise par le logiciel Axis Neuron fourni avec la combinaison: c'est l'opération de motion retargeting. Il faut ensuite manipuler l'avatar en parallèle du contrôle par le mocapteur en vue d'ajuster les adresses scéniques. Enfin, il est nécessaire de gérer sur une interface facilement accessible les événements successifs qui interviennent au fil d'un scénario d'improvisation, et permettre des accès à cette interface par des périphériques extérieurs, les plus utilisés étant un gamepad/joystick ou un contrôleur midi type NanoKontrol2 (fabriqué par Korg). Rémy Sohier, enseignant-chercheur collègue du laboratoire INREV vient renforcer le tandem. Un premier environnement de programmation appelé AKN_Core, directement inspiré de la librairie AKeNe, est prototypé au sein d'UE 4.15 à l'occasion de CW#6 du 26 au 28 juin 2017 avec trois équipes qui expérimentent chacune des configurations différentes de la relation entre mocapteurs, avatars et acteurs masqués.

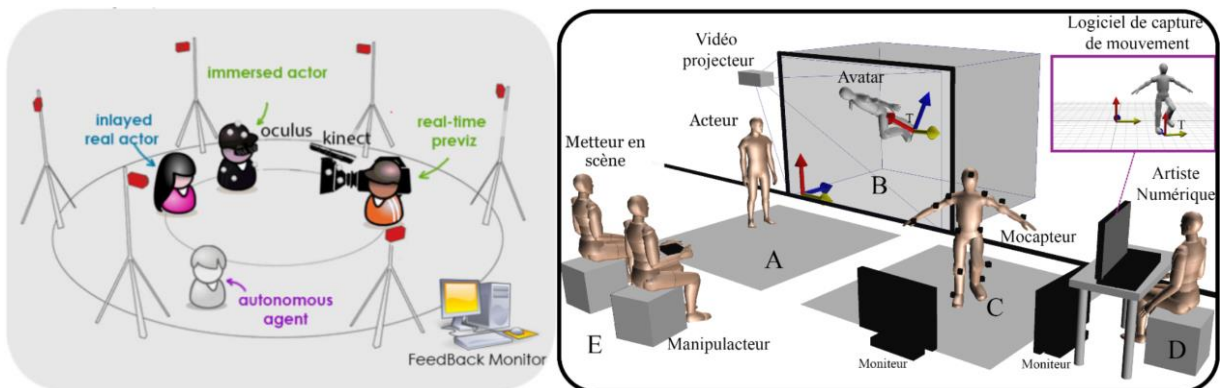

*Figure 3. Gauche : dispositif OUTILNUM.. Droite : dispositif AvatarStaging.*

La volonté de portabilité du dispositif conduit à travailler avec le plugin Perception Neuron d'acquisition des informations d'Axis Neuron, directement intégré à UE4, mais limité concernant l'adaptation à la diversité des squelettes d'avatars utilisés. L'utilisation d'une architecture de *poseable mesh* (cf. note 16) résout une partie des problèmes de motion retargeting et ouvre des perspectives de partage du contrôle d'un avatar par plusieurs mocapteurs et autres périphériques extérieurs. Une architecture de gestion des événements (*cues*) permet d'écrire graphiquement dans un blueprint la succession des actions requises pendant une improvisation. Les outils prototypés par Cédric Plessiet et Rémy Sohier sont directement confiés à Georges Gagneré qui se charge de les configurer selon les besoins de chacune des équipes. Les gamepads sont confiés à des acteurs (appelés « manipulacteurs ») qui assistent les mocapteurs dans le contrôle en rotation et translation des avatars. Le contrôleur NanoKontrol2 sert à déclencher les *cues*. Après trois jours d'expérimentations, de nouveaux besoins émergent, des ajustements s'avèrent nécessaires, les canevas d'improvisation sont modifiés, mais globalement, Perception Neuron et l'environnement AKN_Core laissent présager des résultats publiquement présentables en fin d'année.

---

[15] Georges Gagneré, Cédric Plessiet et Rémy Sohier, « Interconnected virtual space and Theater. Practice as research on theater stage in the era of the network », *in Challenges of the Internet of Things. Technology, Use, Ethics*, dir. Imad Saleh & al., Wiley, 2018.

### 3.2. Un terrain d'apprentissage pour *power user*

Les deux chercheurs du laboratoire INREV travaillent une grande partie de l'été afin de livrer au metteur en scène une nouvelle version d'AKN_Core pour le prochain CW#7, cette fois-ci sous forme de plugin indépendant, facilement adaptable de projet en projet et résolvant une grande partie du cahier des charges établi fin juin[16]. A partir de ce moment, AKN_Core est laissé entre les mains du metteur scène, *power user* (super utilisateur), qui explore progressivement l'architecture des blueprints qu'on lui a confiée, s'efforce d'en comprendre la logique, et se risque à l'ajuster en fonction des besoins du plateau. Le baptême du feu a donc lieu du 4 au 17 octobre 2017. Au fil de chaque répétition, de nouveaux besoins surgissent, de nouvelles configurations des fonctionnalités s'imposent, des améliorations ergonomiques deviennent évidentes. Grâce à l'accessibilité de la programmation graphique par blueprints, Georges Gagneré outrepasse son rôle de *power user* et développe progressivement ses propres outils permettant d'agencer les fonctionnalités d'AKN_Core au sein d'un groupe de blueprints intitulé AKN_Tool. A l'issue de l'atelier, il se construit un prototype de gestion des *cues* pour être encore plus réactif en prévision d'un dernier CW#8 où chaque projet finalise ses expérimentations et les présente au public les 14 et 15 décembre 2017 au Cube, centre de création numérique situé à Issy-Les-Moulineaux.

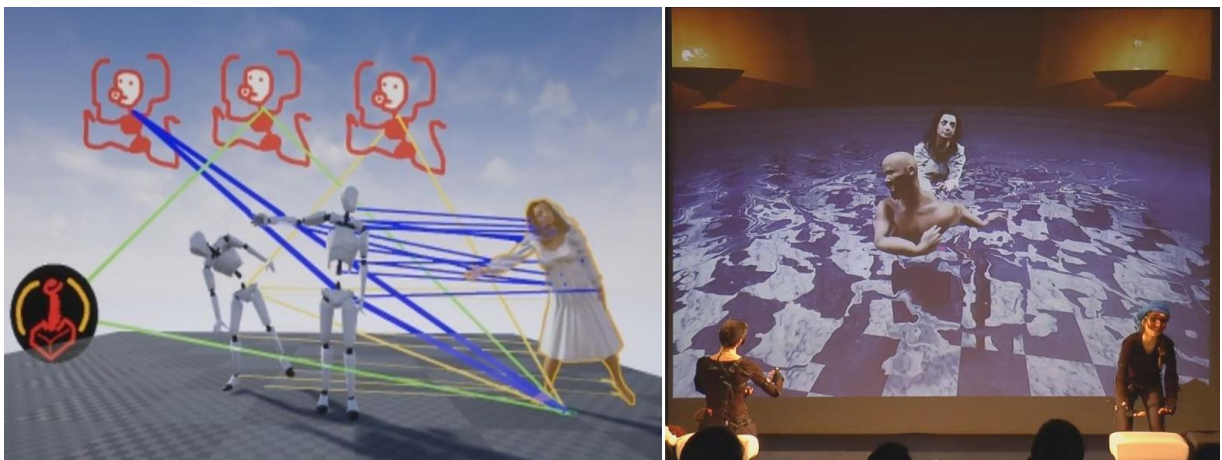

*Figure 4 : Gauche : AKN_Core sous UE4. Droite :* Agammenon Redux *en décembre 2017.*

Les résultats obtenus satisfont à la fois le public spectateur et les artistes réalisateurs. Des enjeux de motion retargeting et de contrôle des avatars avec gamepad n'ont toutefois pas totalement été résolus. Ils proviennent de la construction des avatars et de la manière dont le plugin Perception Neuron pour UE4 est programmé. Le plugin ne fonctionne correctement que pour un type de squelette d'avatar. Les potentialités expressives ouvertes par les outils du jeu vidéo sont très stimulantes, mais les difficultés pour construire une architecture efficace de

---

[16] Cédric Plessiet, Georges Gagneré et Rémy Sohier, « Avatar Staging: an evolution of a real time framework for theater based on an on-set previz technology », *in 2018 Virtual Reality International Conference*, VRIC '18, ACM, 2018, New York, USA.

P a g e 9 | 12

traitement des données, de la capture de mouvement à la projection sur des avatar élaborés, sont élevées, d'autant plus lorsque les utilisateurs sont des acteurs spécialistes de l'expression gestuelle, et que l'on travaille en temps réel, sans possibilité de corriger les animations. La collaboration entre spécialistes du jeu vidéo et du théâtre permet de trouver des solutions aux problèmes soulevés. Le jeu vidéo catalyse ainsi l'émergence de nouvelles formes scéniques.

### 3.3 Refonte des outils : AKN_Regie

A l'issue de *Masques et Avatars*, une des trois équipes conduite par Andy Lavender, enseignant-chercheur britannique, décide de poursuivre l'exploration au sein de son laboratoire de recherche à Warwick University. C'est l'occasion d'une refonte d'AKN_Core sur deux points : comme Andy Lavander souhaite travailler avec des avatars réalistes, Cédric Plessiet propose d'utiliser à nouveau directement les *skeletal meshes*[17] d'UE4 et de faire le motion retargeting des données d'Axis Neuron à nouveau dans Motion Builder, comme cela était le cas lors du 3ème atelier CréaTIC avec l'Optitrack (cf. 1.4.). Il prévoit d'utiliser une brique logicielle d'AKeNe pour établir la relation entre Motion Builder et AKN_Core dans UE4. Par ailleurs, Georges Gagneré souhaite améliorer son module AKN_Tool afin de le rendre plus facilement accessible à des non-spécialistes en programmation blueprint, notamment Tim White, collaborateur d'Andy Lavender, responsable du design numérique de leur projet intitulé *Agamemnon Redux*. Les problèmes de contrôle en rotation et en translation des avatars avec gamepad sont résolus. En parallèle de la refonte du module, un embryon de documentation est mis en place afin de permettre aussi aux étudiants en théâtre de Paris 8 d'utiliser AKN_Core, notamment Anastasiia Ternova, étudiante en master à l'époque, qui accompagne un autre projet étudiant de *Masques et Avatars*. Deux opérateurs utilisent donc de manière autonome AKN_Core et AKN_Tool pour développer leurs idées et approfondir la qualité des interactions entre des avatars numériques et des performers physiques en réalité mixte dans deux dispositifs différents. Les deux expériences se concluent avec succès au mois de mars 2018, à Warwick dans le cadre de l'événement public *Mask and Avatar - Engagement Day* le 23, et à l'université Paris 8 pour le projet étudiant avec cinq représentations dans le cadre de la Semaine des Arts du 26 au 30.

Malgré les développements ingénieux de Cédric Plessiet entre Motion Builder et UE4, les difficultés de motion retargeting résistent. Georges Gagneré décide alors d'inverser les priorités du cahier des charges dans la relation entre théâtre et jeu vidéo. Au lieu de toujours demander l'impossible aux artistes-programmeurs, il rassemble toutes les briques de programmation qui ont déjà fait leur preuve, réécrit sous forme de blueprints l'ensemble des modules nécessaires à l'animation numérique des avatars dans le dispositif AvatarStaging et le nomme AKN_Regie. L'architecture est très proche des précédentes versions d'AKN_Core et AKN_Tool, mais les développements sont simplifiés en utilisant des fonctionnalités spécifiques à la version d'UE 4.18. Ce module AKN_Regie permet de n'utiliser que des avatars compatibles avec les plugins UE4 des périphériques de capture de mouvement généralement fournis par les fabricants. Pour Perception Neuron, il s'agit d'un robot assez gracile, mais sans la dimension réaliste des précédents avatars utilisés par les équipes artistiques. La documentation de l'utilisation d'AKN_Regie est considérablement développée sur un site dédié[18] ce qui permet à toute personne motivée de l'utiliser après avoir téléchargé une version d'UE4. Les usages pédagogiques et artistiques du dispositif AvatarStaging et du module AKN_Regie sont présentés à l'occasion d'un workshop à l'International Conference on Movement Computing en juin 2018[19]. Malgré les contraintes esthétiques sur la nature des avatars, le module est à nouveau utilisé par Tim White pour une troisième instanciation du projet *Agamemnon Redux* à la fondation Cacoyannis d'Athènes le 7 octobre 2018. La résolution des problèmes de motion retargeting permet d'approfondir les enjeux de direction d'avatar[20]. AKN_Regie est aussi utilisé par Anastasiia Ternova pour conduire la recherche-création de son master impliquant entre autres deux mocapteurs et deux manipulacteurs.

---

[17] Un *skeletal mesh* est une version propre à UE4 de l'architecture *poseable mesh* d'animation des avatars. Il permet un contrôle des joints en rotation grâce à un blueprint d'animation.
[18] http://avatarstaging.eu (consulté le 22/01/2020)

**4. Retour à la marionnette virtuelle : CAVOAV (2019)**
**4.1 Apprendre la patience pour finaliser les chantiers en cours**
Début 2019, Georges Gagneré souhaite reprendre sa recherche-création avec les ombravatars qu'il avait suspendue à l'issue de sa troisième résidence *ParOral* à Bordeaux en avril 2016. L'énergie de développement avait été entièrement absorbée par les efforts pour mettre sur pied une plate-forme d'expérimentation suffisamment robuste et manipulable pour obtenir des résultats intéressants à la fois sur les plans de la pédagogie, de la création artistique et de la recherche. Il s'avère que Cédric Plessiet avait repris de son côté des travaux sur la compréhension de la nature des entités virtuelles peuplant les dispositifs d'arts numérique et vidéoludique, et était parvenu à une classification mettant en regard quatre catégories fondamentales : les marionnette, acteur, golem et masque virtuels[21], influencé par sa collaboration récente avec le théâtre. Cette classification permet de comprendre ce qui distingue les êtres numériques artificiels sous l'angle de la décision et du mouvement et suscite l'intérêt du metteur en scène de se confronter à la figure du golem virtuel, figure plus accessible avec les outils actuels d'intelligence artificielle pour les jeux vidéo, que l'acteur virtuel autonome. Le travail sur les ombravatars pourrait en permettre une première approche.

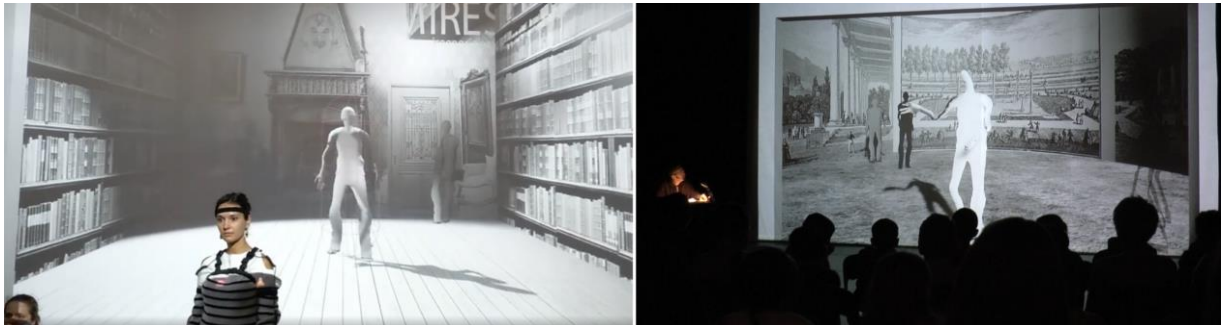

*Figure 5 : Utilisation de CAVOAV en master classe (gauche) et dans* L'Ombre *(droite).*

Cependant, l'architecture de développement d'AKN_Regie limitait la gamme des avatars utilisables, les ombravatars n'en faisaient pas partie. Il s'avère alors qu'Epic Game venait de sortir avec la version UE 4.19 de mars 2018, le plugin LiveLink permettant une simplification de la communication des informations en temps réel entre le moteur de jeu vidéo et d'autres logiciels, dont notamment Motion Builder. Ce que Cédric Plessiet avait prototypé seul dans son laboratoire pour répondre aux besoins des metteurs en scène, Epic Game et ses équipes de développement en sortaient finalement une version directement intégrée à UE4, validant a posteriori les intuitions de l'artiste numérique. LiveLink permet en effet de recourir à nouveau à la puissance de motion retargeting de Motion Builder. À partir de février 2019, c'est l'occasion pour Georges Gagneré d'écrire une nouvelle version d'AKN_Regie dans laquelle n'importe quel avatar peut être contrôlé par tous les périphériques de motion capture fournissant un plugin pour Motion Builder et manipulé ensuite dans UE 4.22.

---

[19] Georges Gagneré et Cédric Plessiet, « Experiencing avatar direction in low cost theatrical mixed reality setup », *in Proceedings of ACM MOCO conference*, Genova, Italy, June 2018.
[20] Georges Gagneré et Cédric Plessiet « Espace virtuel interconnecté et Théâtre (2). Influences sur le jeu scénique », *in Revue : Internet des objets*, Numéro 1, Volume : 3, février 2019, ISSN : 2514-8273, ISTE OpenScience, 2019.
[21] Cédric Plessiet, Georges Gagneré et Rémy Sohier, « A Proposal for the Classification of Virtual Character », *in Proceedings of the 14th International Joint Conference on Computer Vision, Imaging and Computer Graphics Theory and Applications* - Volume 2: HUCAPP, Prague, Czech Republic, 2019, p. 168-174.

**4.2. De *ParOral* à CAVOAV : mettre en scène l'ombre numérique**
Le premier objectif de cette nouvelle version est ainsi de permettre de retravailler avec l'ombravatar utilisée pour *ParOral*. La grande mobilité spatiale offerte au mocapteur par la Perception Neuron change la donne par rapport aux possibilités de direction scénique de l'ombravatar. Il ne s'agit plus de transférer des gestes sur des marionnettes en position fixe, mais d'écrire des interactions complexes entre plusieurs ombravatars habitant des scénographies 3D variées. AKN_Regie permet alors de construire un environnement numérique créatif dédié au travail avec les ombres numériques : le CAstelet Virtuel d'OmbrAVatar (CAVOAV). Une caractéristique de CAVOAV est de permettre d'enregistrer le parcours scénique d'une ombravatar et de le rejouer en pouvant suspendre l'action à des moments clefs[22]. CAVOAV permet de faire aboutir à l'automne 2019 une version complète de la mise en scène de *L'Ombre*, cinq ans après la conception du projet en 2014[23] (cf. fig. 5).

**5. Perspectives**
L'imbrication entre les orientations de développement et les expérimentations pratiques théâtrales qui ont été réalisées entre 2014 et 2019 est riche d'enseignements. Des potentialités offertes par les technologies du jeu vidéo sont laissées de côté pour faire avancer d'autres facettes qui à leur tour sont mises entre parenthèses en attendant que des solutions soient débloquées par de nouvelles avancées. La dimension d'expérimentation en immersion a par exemple été momentanément délaissée à partir de 2017. Cependant, le travail immergé en réalité virtuelle renouvellera les techniques du mocapteur. Il permettra de faire voyager le public de théâtre entre réalité physique et immersion.

L'évolution de Georges Gagneré vers une appropriation progressive mais inéluctable de la programmation d'un moteur de jeux vidéo[24] traduit la place croissante qu'occupent ces derniers dans la réalité mixte des expériences artistiques menées. Il ne s'agit pas seulement de manipuler des images, il faut aussi diriger des avatars et déployer des scénographies 3D aux possibilités infinies. La démarche d'un artiste numérique tel Cédric Plessiet qui déploie des compétences spécifiques pour programmer ses outils de création doit inspirer les artistes de théâtre. Il faut enfin rappeler l'importance de documenter les processus de recherche-création et notamment les méthodes, outils pratiques et technologies qui les rendent possibles, afin d'en faciliter le partage et finalement l'enrichissement par les usages artistiques.

---

[22] Georges Gagneré et Anastasiia Ternova, « A CAstelet in Virtual reality for shadOw AVatar (CAVOAV) », *in ConVRgence (VRIC) Virtual Reality International Conference Proceedings,* dir. Simon Richir, International Journal of Virtual Reality, 2020
[23] *L'Ombre*, d'après Andersen, mise en scène/programmation : Georges Gagneré, jeu : Pavlo Chirva, Eric Jakobiak, Clermont Pithan, musique : Tom Mays, assistanat : A.Ternova. Création en Ukraine (septembre 2019), reprises en France (novembre 2019).
[24] Georges Gagneré, « Du théâtre à l'informatique : bascule dans un nouveau monde », *in Actes des Journées d'Informatique Théâtrale*, février 2020, Grenoble.